\documentclass[a4paper, oneside, twocolumn, notitlepage, 10pt]{extarticle_ecoc} 
\usepackage{ecoc}
\usepackage{circuitikz}
\usepackage{comment}
\usepackage{pgfplots}
\usepackage{amsmath, amssymb, amscd, amsthm, amsfonts,xpatch}
\usepackage{multirow}
\usepackage{pgfplotstable}
\usetikzlibrary{automata,chains,shadows,fit,backgrounds,shapes,arrows,positioning,calc,decorations.pathreplacing}
\usetikzlibrary {arrows.meta,shapes.misc,calc}
\usetikzlibrary{graphs,graphs.standard,quotes,fillbetween}
\pgfplotsset{compat=newest}

\begin{document}
\selectlanguage{english}

\title{Reducing the Error Floor of the Sign-Preserving Min-Sum LDPC Decoder via Message Weighting of Low-Degree Variable Nodes}%
\author{
    Lotte Paulissen, Alex Alvarado, Kaiquan Wu, and Alexios Balatsoukas-Stimming
}

\maketitle              

\begin{strip}
 \begin{author_descr}

Eindhoven University of Technology, 5600MB Eindhoven, The Netherlands,
   \textcolor{blue}{\uline{a.alvarado@tue.nl}}
 \end{author_descr}
\end{strip}

\setstretch{1.1}
\renewcommand\footnotemark{}
\renewcommand\footnoterule{}
\let\thefootnote\relax\footnotetext{978-1-6654-3868-1/21/\$31.00 \textcopyright 2022 IEEE}


\begin{strip}
  \begin{ecoc_abstract}
  Some low-complexity LDPC decoders suffer from error floors. We apply iteration-dependent weights to the degree-3 variable nodes to solve this problem. When the 802.3ca EPON LDPC code is considered, an error floor decrease of more than 3 orders of magnitude is achieved.
  \end{ecoc_abstract}
\end{strip}

\section{Introduction}

Increasing data rates in fiber optical communication systems require powerful forward error correction schemes to provide reliable transmission. High-throughput systems also have stringent power consumption requirements, therefore, hardware-friendly decoders are key importance. The 802.3ca EPON standard \cite{2020802.3ca-2020Networks.} has adopted low-density parity check (LDPC) codes due to their near-capacity performance, but also because LDPC decoders are highly parallelizable.

The belief-propagation (BP) \cite{Richardson2001TheDecoding} algorithm for LDPC decoding offers excellent performance \cite{Chung2012OnLimit} but is complex to implement due to the hyperbolic tangent update rule at the check nodes (CNs). The complexity can be reduced by simplifying the update rule at the CNs, leading for example to the popular min-sum (MS) decoder\cite{Fossorier1999ReducedPropagation}. Recently, the sign-preserving min-sum (SP-MS) decoder was introduced\cite{Cochachin2021Sign-PreservingDecoders} as a low-complexity alternative for the MS decoder. The SP-MS decoder is designed for quantized messages taken from a finite alphabet (with typical alphabet sizes of 4, 8, or 16) and it was shown to perform well for the regular LPDC code with degree-6 variable nodes (VNs) \cite{Cochachin2021Sign-PreservingDecoders} defined in the IEEE 802.3 10G Ethernet standard\cite{2016IEEEEthernet}. Using the SP-MS decoder with small alphabet sizes is potentially very interesting for high-speed optical fiber applications.

In this paper, we study the applicability of very low complexity versions of the SP-MS decoder (alphabet sizes 4 and 8) for the 802.3ca EPON LDPC code. The first contribution of this paper is to show that such decoders result in an error floor for the EPON LDPC code. We also show that the reason for this floor is the degree-3 VNs present in the 802.3ca EPON LDPC code and the difficulties very low complexity versions of the SP-MS decoder has dealing with such low-degree VNs\cite{Lechner2010AnalysisDecoders}. 

The second contribution of this paper is to propose a method to reduce the error floor for the SP-MS decoder. The main idea is to weight up the CN-to-VN messages to allow the decoder to ``escape'' situations where the channel messages dominate the strongly quantized messages in the SP-MS decoder. This idea also means that weights are only required for low-degree VNs, while for high-degree VNs, the incoming messages can make the decoder escape problematic situations.
 
Numerical results for the 802.3ca EPON LDPC code show that the bit error rate (BER) error floor can be reduced by more than three orders of magnitude by applying weights to degree-3 VNs only. These results are based on a hardware-friendly implementation of our algorithm, where all weights are represented using three bits only.

\section{Quantized Decoding}

\begin{figure}[b]
\centering
\setkeys{Gin}{width=1\textwidth}
\includegraphics[width=0.98\linewidth]{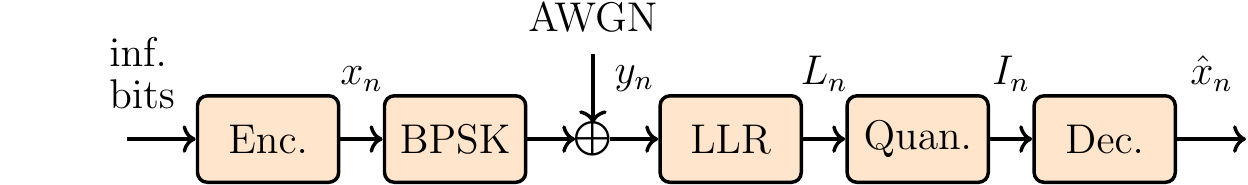}
\vspace{0.5ex}
\caption{System model.}
\label{fig:channel_model}
\end{figure}

As shown in Fig.~\ref{fig:channel_model}, LDPC coded bits $x_n$ ($n = 1,2,\ldots,N$) are modulated using binary-phase shift keying (BPSK) and transmitted over the additive white Gaussian noise (AWGN) channel. At the receiver, log-likelihood ratios (LLRs) $L_n$ are calculated. In this paper, we use 4 bits for the quantization of the LLRs. Using more bits for the LLR quantization does not affect the conclusions in this paper. 
In the quantization process, a constant scaling factor $\alpha$ is used to map the channel LLRs to their corresponding quantized LLRs $I_n$. When required, the scaling $\alpha$ was numerically optimized in the numerical simulations. These quantized LLRs are used as input of the LDPC decoder.

In the LDPC decoder, messages are iteratively exchanged between VNs and CNs in order for the decoder to converge towards a valid codeword. A VN-to-CN message at iteration $\ell$ is denoted as $m_{v_n\rightarrow c_m}^{(\ell)}$, and $m_{c_m\rightarrow v_n}^{(\ell)}$ denotes a CN-to-VN message at iteration $\ell$. All CNs that are connected to VN $n$ except for CN $m$ are denoted by $c_{m}^{\text{ext}}$. At the VNs of the BP decoder, we add the likelihoods from all neighbouring CNs to the channel message. For the unquantized BP decoder, this can be defined as
\begin{equation}\label{VN.1}
    m_{v_n \rightarrow c_m}^{(\ell+1)} = L_n+ \sum_{c_{m}^{\text{ext}}} m_{c\rightarrow v_n}^{(\ell)}.
\end{equation}

We target low-complexity decoders where the messages in the LDPC decoder are strongly quantized. We consider $q = \{2,3,4\}$ precision bits, with alphabet sizes $4, 8$ and $16$ resp. When the messages in the decoder are quantized to $2$ or $3$ precision bits, there is a mismatch between the channel LLRs (quantized using $4$ bits) and incoming messages in the VN update rule (see \eqref{VN.1}). As we will show below, this causes an error floor.

In this paper, we propose to modify the SP-MS  decoder\cite{Cochachin2021Sign-PreservingDecoders} by adding an iteration dependent weight $w_n^{(\ell)}$. The modified SP-MS update rule is 
\begin{equation}\label{VN.2}
    m_{v_n \rightarrow c_m}^{(\ell+1)} = \Psi \Bigg(I_n+ w_n^{(\ell)}\Big(\frac{\mu_{v_n \rightarrow c_m}^{(\ell)}}{2}+\sum_{\scriptscriptstyle c_{m}^{\text{ext}}} m_{c\rightarrow v_n}^{(\ell)}\Big)\Bigg),
\end{equation}
which coincides with the original SP-MS decoder when $w_n^{(\ell)}=1$ for $n=1,2,..,N$ and $\forall \ell$.

In \eqref{VN.2}, $\mu_{v_n \rightarrow c_m}^{(\ell)}$ is the sign-preserving factor to ensure the VN update rule never generates an erased message, which is a message that carries no information, and therefore, does not contribute to the convergence of the decoder. Additionally, quantized decoders require a function $\Psi (m_s) = \left(\operatorname{sign} (m_s), \mathcal{S}\left( \max \left( \lfloor\left| m_s\right|\rfloor-\varphi_v,0\right),2^{q-1}-1\right)\right)$, to saturate the outgoing messages back to the specified message alphabet. $\mathcal{S}(\cdot)$ is a saturation function clipping the message $m_s$ to $2^{q-1}-1$ when its magnitude is larger than $2^{q-1}-1$, and $\varphi_v$ is an offset factor dependent on the magnitude of the message\cite{Cochachin2021Sign-PreservingDecoders}. 

In this paper, we consider the IEEE 802.3ca EPON irregular LDPC code\cite{2020802.3ca-2020Networks.} with $N=17664$, $R=0.826$, and with VN degree distribution polynomial 
{\small{\begin{equation}\label{code}
\lambda(X)=\frac{12800}{17664}X^2+\frac{4352}{17664}X^5+\frac{256}{17664}X^{10}+\frac{256}{17664}X^{11}.
\end{equation}}}
Throughout this paper, we always consider a maximum of $12$ decoding iterations. For all simulation results, a minimum of 500 frames are sent with at least 30 frame errors occurring for each signal-to-noise ratio (SNR) point.

\begin{figure}[t]
\centering
\setkeys{Gin}{width=1\textwidth}
\includegraphics[width=0.98\linewidth]{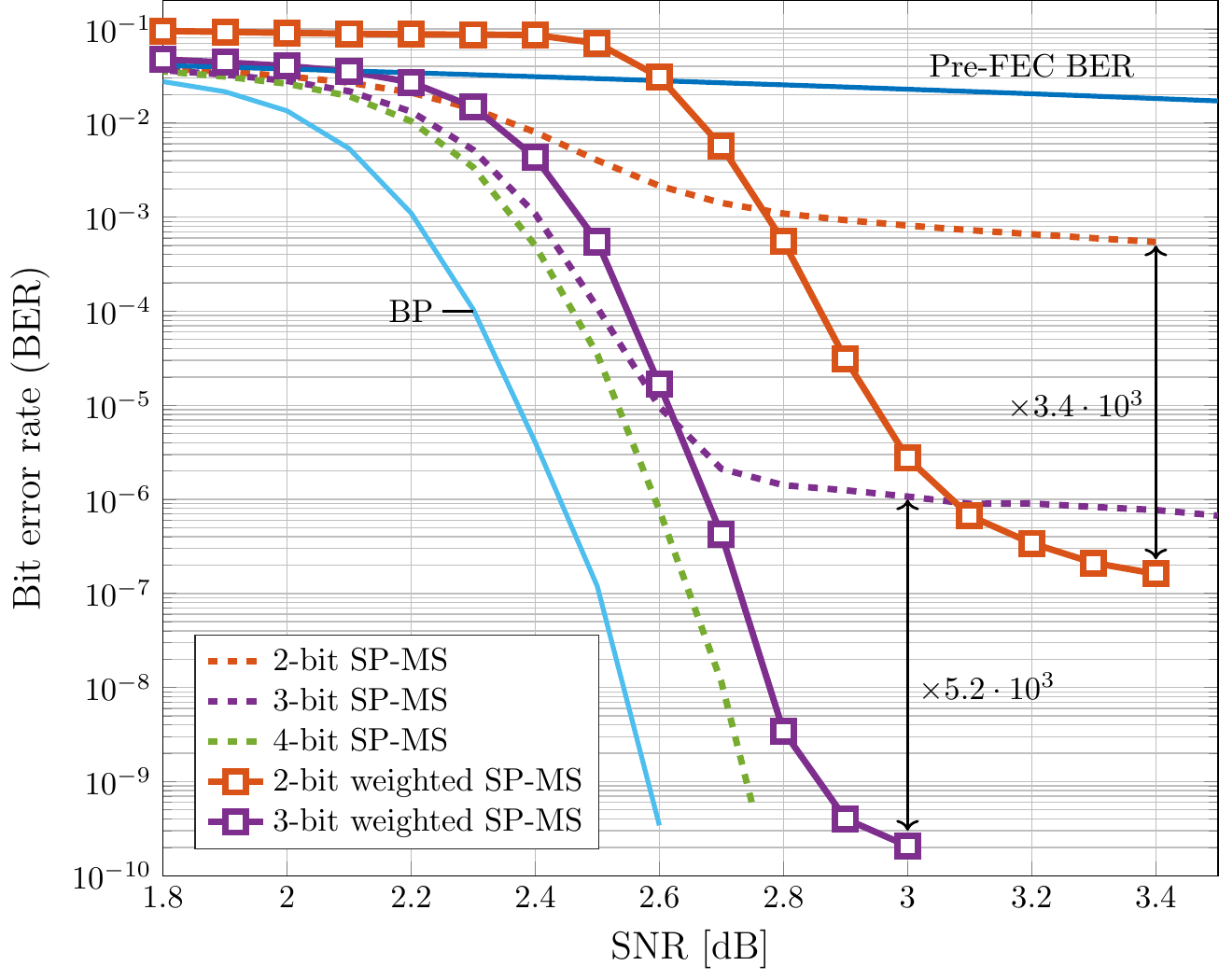}
\includegraphics[width=0.98\linewidth]{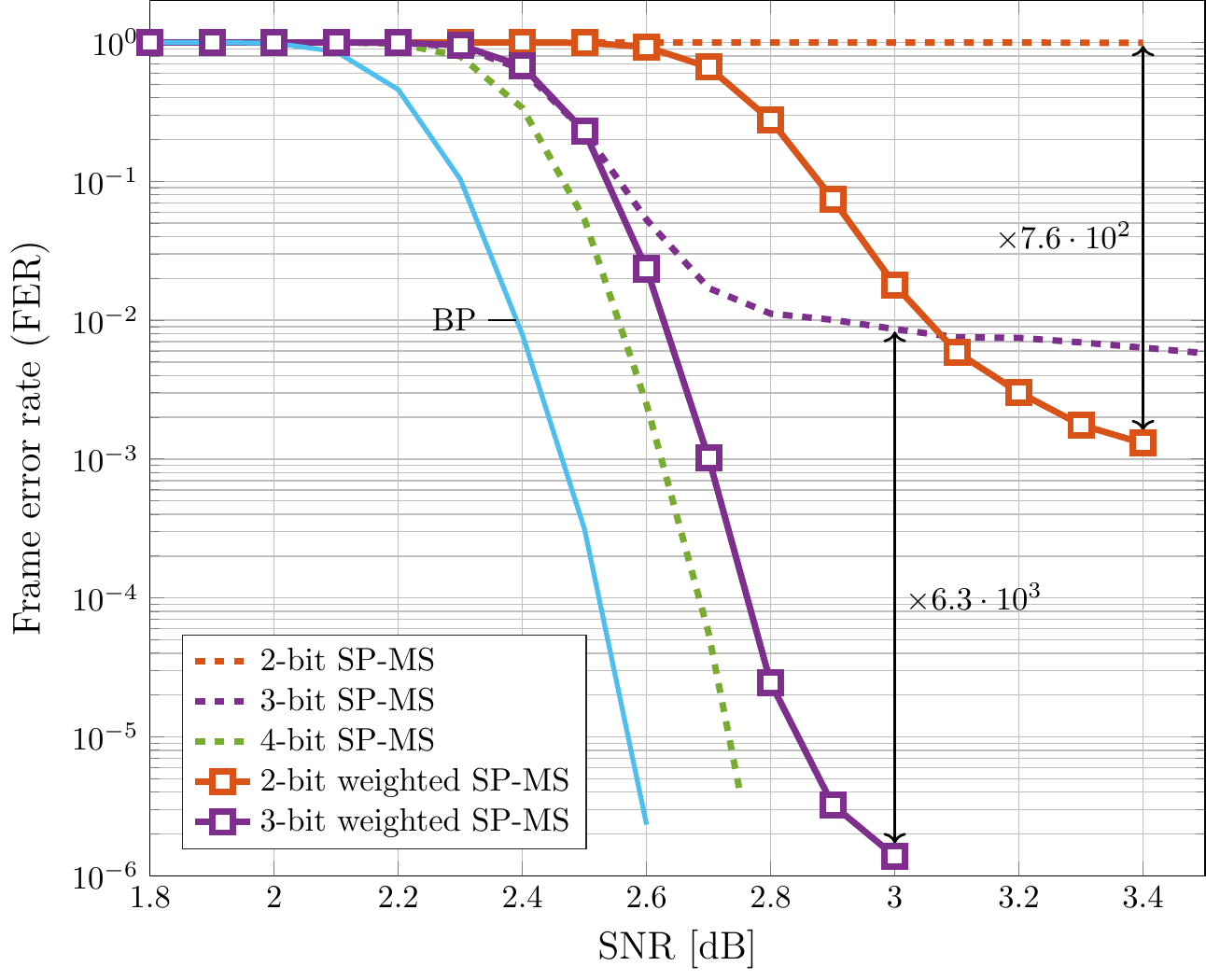}
\caption{BER (top) and FER (bottom) performance of the $q$-bit SP-MS decoder (dashed lines) and the proposed $q$-bit degree-3 weighted SP-MS decoder (solid lines with squares).}
\label{Fig:Results}
\end{figure}

Fig.~\ref{Fig:Results} shows the BER and frame error rate (FER) performance of the SP-MS decoder (using \eqref{VN.2} with $w_n^{(\ell)}=1$) with $q = \{2,3,4\}$ (dashed lines). These results were obtained using $\alpha=0.75,0.95,1.15$ for $q=2,3,4$, resp. As reference, we also show the pre-FEC BER and the performance of the BP decoder using \eqref{VN.1} (solid lines). These results indicate that SP-MS with $q=4$ (dashed green) offers performance close to that of the BP decoder, but they also show an unacceptably high error floor for the cases $q =2$ and $q =3$ (BER $\approx 10^{-3}$ and $\approx 10^{-6}$, resp). Similar conclusions can be drawn from the FER results. We will show in the next section that by adjusting $w_n^{(\ell)}$ in \eqref{VN.2}, the error floor can be significantly lowered.

\section{Weighted SP-MS Decoder}

\begin{table*}[t]
    \centering
    \caption{Optimized weights $w_n^{(\ell)}$ for 802.3ca LDPC code.}{\footnotesize
    \begin{tabular}{c|c|c|c|c|c|c|c|c|c|c|c|c}
         \hline 
         
         \hline
         $q$ & $w_n^{(0)}$ & $w_n^{(1)}$ & $w_n^{(2)}$ & $w_n^{(3)}$ & $w_n^{(4)}$ & $w_n^{(5)}$ & $w_n^{(6)}$ & $w_n^{(7)}$ & $w_n^{(8)}$ & $w_n^{(9)}$ & $w_n^{(10)}$ & $w_n^{(11)}$ \\
         \hline 
         
         \hline
         $2$  & $1.0$ & $1.0$ & $1.0$ & $1.0$ & $1.5$ & $1.75$ & $1.75$ & $1.75$ & $1.75$ & $2.5$ & $3.0$ & $3.0$ \\ \hline
         $3$ & $1.0$ & $1.0$ & $1.0$ & $1.0$ & $1.25$ & $1.25$ & $1.25$ & $1.25$ & $1.5$ & $1.5$ & $1.5$ & $1.5$ \\ 
         \hline
        
        \hline 
    \end{tabular}

    }
    \label{tab:weight}
\end{table*}

In this section, we propose to modify the SP-MS decoder by introducing the coefficient $w_n^{(\ell)}$ in \eqref{VN.2}.
The rationale behind this is as follows. For the BP decoder, by performing the update rules and exchanging messages iteratively, more bits are likely to be correct at each iteration. Therefore the importance of the incoming messages increases compared to the channel LLRs as the iteration number grows. The introduced weighting factor is used to mimic this evolution and to adjust the importance of the magnitude of the incoming messages compared to the channel LLRs in the VN and tentative update rule. This is realized by applying an iteration-dependent weighting factor to the incoming messages.

As mentioned before, for low values of $q$, there is a mismatch between the alphabets of the channel LLRs and the messages inside the decoder. We conjecture that the decoder is in general able to overcome this mismatch for VNs with a large number of incoming messages. However, the decoder might be unable to recover from bad channel LLRs for low-degree VNs. In what follows we will show that this is indeed the case and that the error floors in Fig.~\ref{Fig:Results} are due to the degree-3 VNs.

Based on the observation that the LDPC code under consideration has a large number of degree-3 VNs ($\approx 72$~\% of the total, as shown in \eqref{code}), we propose to apply weights $w_n^{(\ell)}\neq 1$ to degree-3 VNs so that the mismatched alphabets can be taken care of. This approach does not only allow us to prove our conjecture but also to reduce complexity as not all VNs need to be weighted. Furthermore, using this approach, the LLR quantization coefficient $\alpha$ does not need to be optimized. For simplicity, in our results we use the same $\alpha$ values used for the SP-MS decoder.

The weights need to be chosen carefully, because suboptimal weights can significantly decrease the performance of the decoder. The weight vectors are determined by selecting the one resulting in the lowest BER/FER performance from a set of random candidate weight vectors. Furthermore, to simplify parameter optimization, avoid overestimation, and to follow the rationale explained above, the weights are constrained such that $0<w_n^{(\ell)}\leq w_n^{(\ell+1)}$. 


The modified SP-MS update rule in~\eqref{VN.2} requires a multiplication operation. The hardware implementation of a multiplication has a high complexity when compared to the remaining operations required by the SP-MS decoder (simple additions and comparisons). Moreover, supporting optimum weights requires a large number of quantization bits, which further complicates the multiplication operation. We thus propose a hardware-friendly implementation of the modified SP-MS decoder, where the weights $w^{(\ell)}_n$ are limited to a sum of up to three powers of two. The obtained results are reported in Tab.~\ref{tab:weight} for $q = \{2,3\}$. 

BER and FER results using the weights in Tab.~\ref{tab:weight} are shown in Fig.~\ref{Fig:Results} (solid lines with squares). By applying iteration-dependent weights to degree-3 VNs quantized to only 3 bits, the error floor can be reduced by more than three orders of magnitude. The weights reported in Tab.~\ref{tab:weight} were optimized for an SNR of $2.8$~dB and $3.1$~dB for $q=3$ and $q=2$, resp. For the first four iterations, the obtained weights $w_n^{(\ell)}$ are equal to one. Therefore, we only require shift-and-add operations for the remaining 8 iterations. 


\section{Conclusions}
In this paper, we have shown that an error floor occurs when we have mismatched alphabets between channel LLRs and incoming messages for the SP-MS decoder. The error floor occurs due to degree-3 VNs, which are the majority of the VNs present in the 802.3ca EPON LDPC code. We proposed a method to reduce this error floor by applying iteration-dependent weights to the incoming messages of degree-3 VNs. Our hardware friendly implementation of the modified SP-MS decoder is able to reduce the error floor by more than three orders of magnitude. This method is a general method that next to the SP-MS decoder, can be used for all low-complexity decoders that suffer an error floor due to mismatched alphabets. 
Further research should be conducted to identify why we observe an error floor at lower BER in order to design a low-complexity LDPC decoder that can achieve error floor-free performance close to the BP decoder. 

\vspace{2ex}

\scriptsize
{
\noindent \textbf{Acknowledgements:}~
The work of K.~Wu and A.~Alvarado has received funding from the Netherlands Organisation for Scientific Research via the VIDI Grant ICONIC (project number 15685). The work of A. Alvarado has also received funding from the European Research Council (ERC) under the European Union's Horizon 2020 research and innovation programme (project 963945).}


\printbibliography
 
\vspace{-4mm}

\end{document}